\documentclass[prl,floatfix,twocolumn]{revtex4}

\def\Ipeak{I_{\mathrm{peak}}}
\def\Leff{L_{\mathrm{eff}}}
\def\Epump{\mathbf{E}_{\mathrm{pump}}}
\def\Eprobe{\mathbf{E}_{\mathrm{probe}}}
\def\tpump{t_{\mathrm{pump}}}
\def\xpump{x_{\mathrm{pump}}}

\usepackage{amsmath}
\usepackage{graphicx}
\usepackage{hyperref}

\begin{document}
\title{Optical nonlinearity in Ar and N$_2$ near the ionization threshold}
\author{J. K. Wahlstrand}
\affiliation{Institute for Research in Electronics and Applied Physics, University of Maryland, College Park, MD 20742 USA}
\author{Y.-H. Cheng}
\affiliation{Institute for Research in Electronics and Applied Physics, University of Maryland, College Park, MD 20742 USA}
\author{Y.-H. Chen}
\affiliation{Institute for Research in Electronics and Applied Physics, University of Maryland, College Park, MD 20742 USA}
\author{H. M. Milchberg}
\affiliation{Institute for Research in Electronics and Applied Physics, University of Maryland, College Park, MD 20742 USA}

\begin{abstract}
We directly measure the nonlinear optical response in argon and nitrogen in a thin gas target to laser intensities near the ionization threshold.
No instantaneous negative nonlinear refractive index is observed, nor is saturation, in contrast with a previous measurement [Loriot \emph{et al.}, Opt. Express \textbf{17}, 13429 (2009)] and calculations [Br\'ee \emph{et al.}, Phys. Rev. Lett. \textbf{106}, 183902 (2011)].
In addition, we are able to cleanly separate the instantaneous and rotational components of the nonlinear response in nitrogen.
In both Ar and N$_2$, the peak instantaneous index response scales linearly with the laser intensity until the point of ionization, whereupon the response turns abruptly negative and $\sim$constant, consistent with plasma generation.
\end{abstract}

\maketitle

The optical Kerr effect, the intensity-dependent refractive index experienced by an optical pulse in a transparent medium, plays an important role in phenomena from nonlinear propagation in optical fibers \cite{agrawal_nonlinear_2006} to mode-locking in pulsed lasers \cite{haus_mode-locking_2000} to filamentary propagation in condensed media and the atmosphere \cite{couairon_femtosecond_2007}.
A recent transient birefringence measurement in the components of air reported by Loriot et al.~\cite{loriot_measurement_2009} purported to show that the optical Kerr effect saturates and then becomes negative for intensities greater than 26 TW/cm$^2$.
A strong higher-order Kerr effect, with a crossover from positive to negative nonlinear index at intensities well below the ionization threshold, would have a huge impact on the nonlinear optics of transparent media, and has inspired theoretical works predicting plasma-free light filamentation \cite{bejot_higher-order_2010} and exotic new effects in light propagation \cite{novoa_fermionic_2010}.
It would overturn the picture most have of the mechanism behind long-range filamentary propagation of intense ultrashort pulses -- as arising from an interplay between self-focusing due to the positive optical nonlinearity from bound electrons and defocusing due to the plasma generated by ionization.
The existence of a higher-order Kerr effect would also have implications for the general nonlinear susceptibility in transparent media \cite{ettoumi_generalized_2010,kasparian_arbitrary-order_2010}, including harmonic generation \cite{kolesik_femtosecond_2010,bjot_higher-order_2011,ariunbold_2011}.

Subsequent experimental studies of light filaments \cite{chen_direct_2010,kosareva_arrest_2011,polynkin_experimental_2011,kolesik_higher-order_2010} have not supported the higher-order Kerr model, with one exception \cite{bejot_transition_2011}.
One measurement \cite{chen_direct_2010} found that the electron density was two orders of magnitude higher than predicted by a calculation including higher-order nonlinearities, but agreed with a simulation based on plasma defocusing alone \cite{bejot_higher-order_2010}.
A physical mechanism for the saturation and negative response was proposed based on the nonlinear response near the threshold of ionization \cite{teleki_microscopic_2010,bree_saturation_2011}.
What is missing from this debate is a direct measurement of the nonlinearity that corroborates or refutes the intensity dependence observed by Loriot \emph{et al}.
Here, we describe such a measurement in Ar and N$_2$ using spectral interferometry.
We find no saturation and no negative instantaneous nonlinear phase, in contrast to the original experiment \cite{loriot_measurement_2009}.

The technique we use, single-shot supercontinuum spectral interferometry \cite{kim_single-shot_2002-1}, provides a single-shot measurement of the transverse space- and time-dependent phase shift of a chirped probe pulse due to a transient nonlinearity induced by a short pump pulse.
The time resolution is given by the probe bandwidth (in this experiment $\sim$15 fs) and the transverse spatial resolution here is 3 $\mu$m.
Loriot \emph{et al.}~used a non-spatially resolved multi-shot technique limited in time resolution by the probe duration of $\sim$90 fs \cite{loriot_measurement_2009}.
They measured the transient birefringence and inferred the higher-order Kerr coefficients from the tensorial symmetry of the nonlinear susceptibilities $\chi^{(5)}$, $\chi^{(7)}$, etc.~\cite{loriot_measurement_2009,stegeman_nonlinear_2011}.
In contrast, we can measure parallel and perpendicular components of the nonlinear response independently.
Previously, supercontinuum spectral interferometry was used with 110 fs pump pulses to study the nonlinear response of air consituents \cite{chen_measurement_2007}, and no sign reversal of the nonlinear index was observed up to intensities where ionization occurs.
However, the use of a gas cell complicates the interpretation of the experiment when the response is highly nonlinear \cite{kim_single-shot_2002}.
Also, at the high pressures used ($>3$ atm) plasma-induced refraction limits the peak intensity.
Here, we measure the nonlinear response of Ar and N$_2$ using a 38 fs pump pulse and a 2 mm thick gas target.

\begin{figure}
\center{\includegraphics[width=8cm]{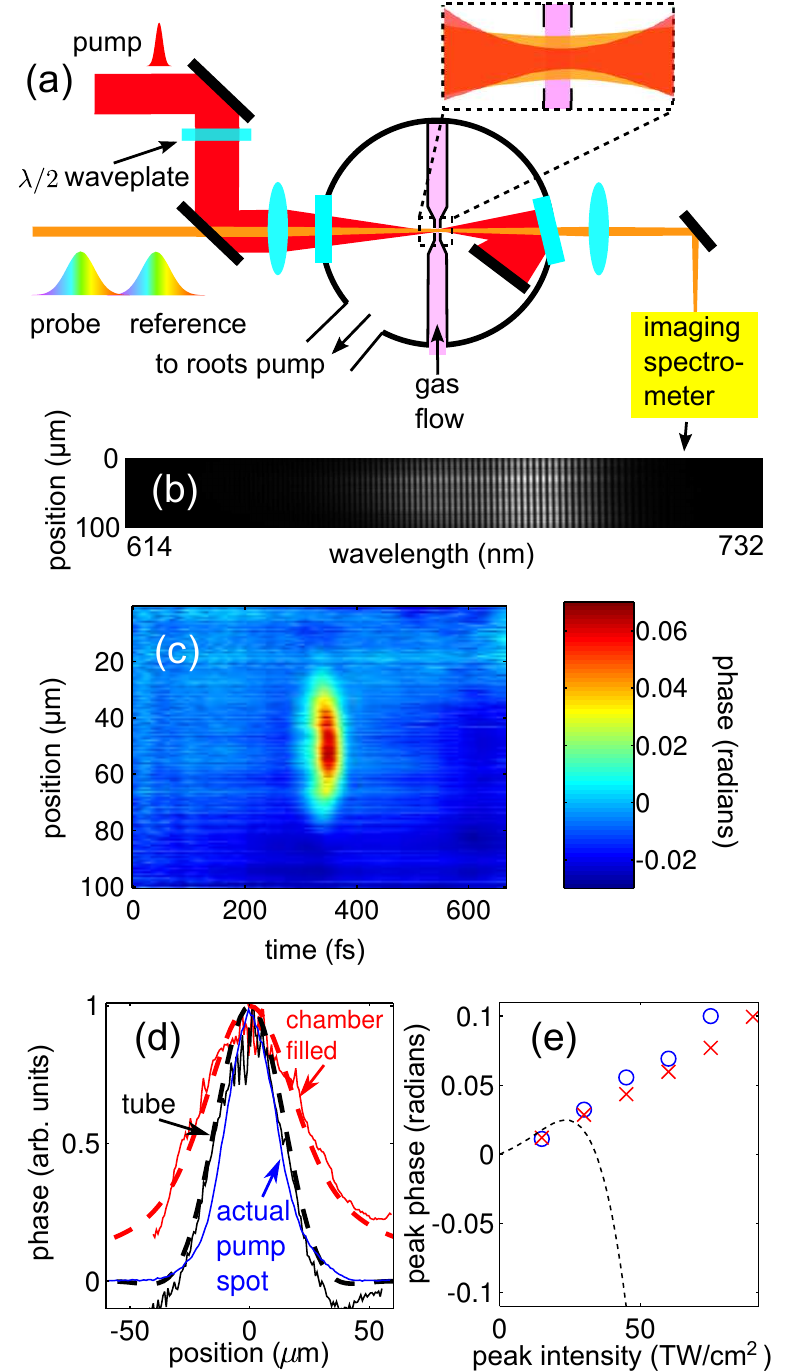}}
\caption{
Experimental apparatus and results.
(a) Simplified experimental setup diagram showing the pump, probe, and reference pulses focused on a flowing gas target in a vacuum chamber.
(b) Example interferogram.
(c) Map of phase versus time and transverse dimension for Ar at $\Ipeak = 60$ TW/cm$^2$, extracted as described in \cite{kim_single-shot_2002-1}.
(d) Lineout of the pump spot, comparing the signal using the drilled flow tube to the signal with a backfilled chamber.
The true pump spot is also shown for comparison.
The dashed lines show the results of propagation simulations \cite{kim_single-shot_2002} for the case of a backfilled chamber and for an interaction length of 2 mm.
(e) The measured peak phase shift for $\Epump \parallel \Eprobe$ as a function of peak intensity for Ar ($\times$) and N$_2$ ($\circ$).
The calculated peak phase shift in Ar using the higher-order Kerr coefficients given in \cite{loriot_measurement_2009} is shown as a dashed line.
}
\label{expt}
\end{figure}

Figure \ref{expt} shows the experimental setup and data.
Detailed descriptions of the experimental technique have been given previously \cite{kim_single-shot_2002-1,chen_measurement_2007}.
The laser is a 1 kHz repetition rate Ti:sapphire amplifier producing 38 fs full width at half maximum (FWHM), 3.5 mJ pulses centered at 800 nm.
Roughly 700 $\mu$J of the laser output is used to generate supercontinuum covering 640-720 nm in a gas cell (not shown) filled with 1-2 atm of Ar; the fundamental is rejected using a dichroic mirror.
The supercontinuum is linearly polarized.
Probe and reference pulses, separated by 1.4 ps, are generated using a Michelson interferometer and chirped so that the group delay dispersion is 1950 fs$^2$, and then the beam is spatially filtered with a 100 $\mu$m pinhole.
The pump power is attenuated using a waveplate and thin film polarizers so that the pulse energy is continuously adjustable from 5-100 $\mu$J.
The pump beam is then expanded with a telescope, and a $\lambda/2$ waveplate allows rotation of the pump polarization.
The pump and supercontinuum are combined using a dichroic mirror; at this point the pump beam is about 5 times wider than the probe/reference beam so that the probe spot overfills the pump spot in the interaction region.

A well-defined interaction length, ideally comparable to or shorter than the Rayleigh range of the pump beam, minimizes spatial and temporal distortions due to axial variation in the pump intensity \cite{kim_single-shot_2002}.
To achieve this, we use a thin gas target inside a vacuum chamber, shown in Fig.~\ref{expt}a.
The beams are focused using a lens of focal length 40 cm, and enter the vacuum chamber through a 5 mm thick fused silica window, propagating 30 cm through vacuum to the target.
The pump beam waist, measured by imaging the focus onto a CCD camera, is 22 $\mu$m FWHM.
The target is a copper gas flow tube with a flattened section through which a 120 $\mu$m diameter hole was laser drilled to allow the pump and probe to pass through.
The space between the inner tube walls is 1 mm, with a wall thickness of 0.5 mm.
A roots pump keeps the background pressure of the chamber at 400 mTorr, and the local gas density drops quickly enough away from the tube that the probe phase shift is dominated by the gas between the holes.
The pump beam is rejected at the exit of the vacuum chamber using a dichroic mirror that is also used as a window.
In the data shown here, we have subtracted a very small background signal due to cross phase modulation in the entrance window.

The central plane of the gas tube is imaged onto the entrance slit of an imaging spectrometer.
Interference fringes appear in the spectrum because of the time delay between the probe and reference pulses.
An example spectral interferogram for Ar at 60 TW/cm$^2$ pump vacuum intensity is shown in Fig.~\ref{expt}b.
The intensity values given here are calculated from the average power and the measured beam profile -- we estimate an uncertainty of 20\%.
The pump pulse, centered at time $t=\tpump$ and transverse dimension $x=\xpump$, causes a phase shift in the probe pulse $\Delta \phi (x,t)$ (too small to produce fringe shifts visible by eye in Fig.~\ref{expt}b).
This causes a change in the spectral phase and amplitude of the probe beam \cite{kim_single-shot_2002-1}.
The spectral phase is found by Fourier analysis of the interference fringes, and the change in amplitude is also found from the interferogram.
The final piece of information required is the spectral phase of the reference pulse, which is, to an excellent approximation, quadratic and proportional to the group delay dispersion \cite{kim_single-shot_2002-1}.
The extracted time domain phase shift $\Delta \phi (x,t)$ of the probe is shown in Fig.~\ref{expt}c.
The signal-to-noise ratio is considerably improved by summing multiple interferograms before performing the phase extraction \cite{chen_measurement_2007}; in all of the data presented here, 300 interferograms were summed at each power and polarization.

The bound electron optical nonlinearity in Ar is instantaneous to a very good approximation because the energy of the lowest electronic excitation is 15 eV, far greater than the photon energy 1.5 eV.
The ordinary instantaneous Kerr effect is linear in the intensity $I$:  the refractive index is of the form $n=n_0+n_2 I$, where $n_0$ is the index of refraction and $n_2$ is the Kerr coefficient, and thus $\Delta \phi (x,t) \propto I_{\mathrm{pump}}(x,t)$.
The measured FWHM in Fig.~\ref{expt}c of 38 fs matches an autocorrelation measurement of the pump pulse.
The spatial profile lineout, shown in Fig.~\ref{expt}d, agrees well with the pump spot profile.
This confirms that the intensity profile in the thin gas target is the same as the vacuum profile.
The deleterious effect of excessive interaction length on the width of the response is illustrated in Fig.~\ref{expt}d.
A lineout of the measured phase shift along $x$ at $t=\tpump$ is shown with the chamber backfilled with Ar -- note the wider profile compared to the flow tube case.
Also shown are simulations of $\Delta\phi(x,\tpump)$ using the beam propagation method \cite{kim_single-shot_2002}; with the simulation we obtain an effective pump-gas interaction length of $\sim$2 mm, in agreement with the tube geometry.
Because of the short interaction length, intensity clamping \cite{couairon_femtosecond_2007} does not affect the intensity profiles.

The peak phase shift measured in the experiment is plotted as a function of peak intensity in Fig.~\ref{expt}e.
We find a very linear dependence for both Ar and N$_2$, and nearly the same Kerr coefficient, which is consistent with other experiments \cite{marceau_femtosecond_2010}.
For peak intensity $\Ipeak=60$ TW/cm$^2$, we measure a peak phase shift in Ar of 0.059 radians.
For a medium with an effective interaction length $\Leff$, $\Delta \phi_{\mathrm{peak}} = 4\pi \Leff n_2 \Ipeak/\lambda$ (note the extra factor of 2 because we measure cross phase modulation).
Using the literature value for Ar, $n_2=9.8\times 10^{-20}$ cm$^2$/W \cite{lehmeier_nonresonant_1985,marceau_femtosecond_2010} at 1 atm, and using $\Leff \approx 2$ mm, we estimate an average pressure in the interaction region of 0.3 atm.
A higher-order Kerr effect would add terms of the form $n_{2m} I^{m}(t)$, where $m>1$ \cite{loriot_measurement_2009}.
No negative instantaneous phase is observed at any intensity in Ar or N$_2$, nor do we see evidence of saturation \cite{bree_saturation_2011}, in disagreement with the results of Loriot \emph{et al.}~\cite{loriot_measurement_2009}.
A simulation of the phase shift expected using the coefficients reported in \cite{loriot_measurement_2009} is shown as a dashed line in Fig.~\ref{expt}e.
The difference is stark and well outside any error in our experiment we can conceive of.
We have also performed the same experiment in Ar using a probe pulse whose spectrum overlaps the pump pulse, with orthogonal polarization so that the pump light could be rejected by a polarizer before the spectrometer (see the appendix).
We have studied the possible origin of the results obtained in \cite{loriot_measurement_2009} and have found that the interference of pump and probe pulses of the same wavelength can produce a plasma grating which gives rise to an effective birefringence \cite{wahlstrand_effect_2011}.

\begin{figure}
\center{\includegraphics[width=8.5cm]{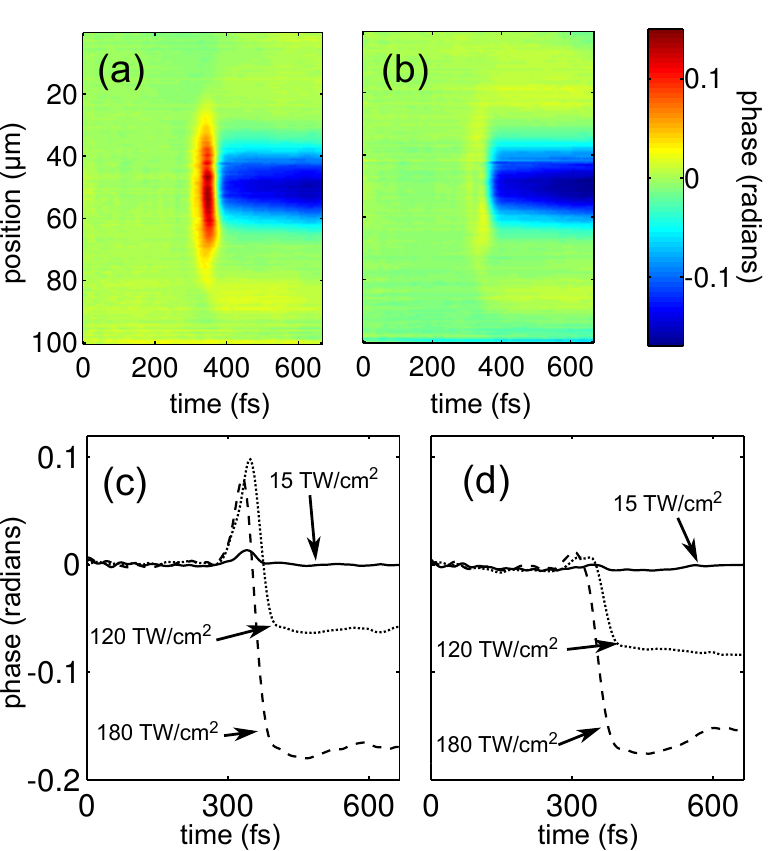}}
\caption{
Experimental data for Ar.
(a,c) $\Epump \parallel \Eprobe$.
(b,d) $\Epump \perp \Eprobe$.
(a,b) Extracted phase $\Delta \phi(x,t)$ at 150 TW/cm$^2$ vacuum pump intensity, showing the positive electronic Kerr effect signal at $t=\tpump$ and the negative plasma signal at later time delays.
(c,d) Lineouts $\Delta \phi(\xpump,t)$ as a function of pump intensity.
Within error, the plasma response is constant in time after the pump has passed; we observe increased noise at the edges of the time window due to the decreased magnitude of the probe/reference spectra on the wings.
The curves at 120 TW/cm$^2$ and 180 TW/cm$^2$ give electron densities $N_e = 7 \times 10^{15}$ cm$^{-3}$ and $N_e = 2 \times 10^{16}$ cm$^{-3}$ respectively.
}
\label{argon}
\end{figure}

Increasing the pump intensity beyond the level of Fig.~\ref{expt} requires careful consideration of increased supercontinuum generation by the pump itself.
Because the pump and probe paths are not phase stable with respect to one another, spectral fringes between the pump supercontinuum and the reference pulse average out when many interferograms are summed.
So the pump supercontinuum does not cause significant data distortion until it saturates the CCD camera, which occurs at intensities higher than 200 TW/cm$^2$, well beyond the ionization threshold of $\sim$100 TW/cm$^2$ \cite{larochelle_non-sequential_1998}.
Maps showing $\Delta \phi(x,t)$ at high intensity in Ar are shown in Fig.~\ref{argon}.
At high intensities we observe an additional response due to the plasma generated by ionization \cite{kosareva_article_1997}.
The plasma produces a negative index contribution $\Delta n_{\mathrm{plasma}}=-N_e/(2N_{cr})$, where $N_e$ is the electron density and $N_{cr}$ is the critical density.
The plasma densities measured are consistent with calculations using Ammosov-Delone-Krainov (ADK) rates \cite{ammosov}.
Plots of $\Delta \phi(\xpump,t)$ as a function of pump intensity are shown in Fig.~\ref{argon}c ($\Epump \parallel \Eprobe$) and Fig.~\ref{argon}d ($\Epump \perp \Eprobe$).
Note that the plasma contribution seen in Fig.~\ref{argon} is highly characteristic:
(1) its onset at higher intensity (180 TW/cm$^2$ compared to 120 TW/cm$^2$) increasingly dominates the instantaneous Kerr response at the back of the pulse; the residual positive Kerr peak appears to move forward in time,
(2) unlike the Kerr response, the plasma-induced phase shift is probe polarization independent, and
(3) after generation, the plasma response is long-lived on the time scale of this measurement, owing to recombination timescales of order $\sim$100 ps.

\begin{figure}
\center{\includegraphics[width=8.5cm]{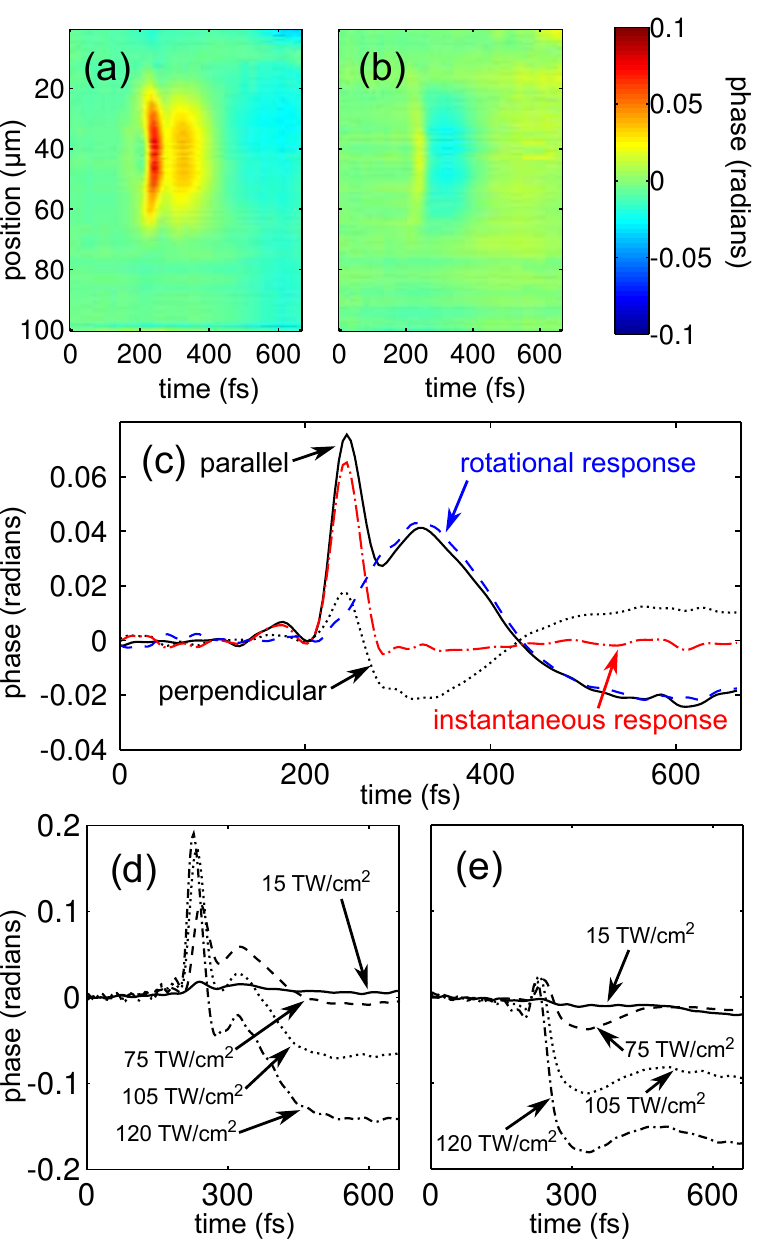}}
\caption{Experimental data for N$_2$.
(a,d) $\Epump \parallel \Eprobe$.
(b,e) $\Epump \perp \Eprobe$.
(a,b) Extracted phase $\Delta \phi(x,t)$ at 60 TW/cm$^2$ pump intensity, showing the positive instantaneous Kerr effect (coincident with the pump) and the rotational response at later time delays.
(c) Lineouts $\Delta \phi(\xpump,t)$ for the data shown in (a),(b).
The decomposition of the signal into instantaneous and rotational components, as described in the text, is also shown.
(d,e) Lineouts $\Delta \phi(\xpump,t)$ as a function of pump intensity.
}
\label{nitrogen}
\end{figure}

In N$_2$, the optical Kerr response has an additional contribution from the transient alignment of the molecules in the strong optical field \cite{nibbering_determination_1997,chen_measurement_2007,chen_single-shot_2007}.
Results for N$_2$ are shown in Fig.~\ref{nitrogen}; $\Delta \phi(x,t)$ is shown at low pump intensity for parallel and perpendicular polarization in Fig.~\ref{nitrogen}ab.
The index change is $\Delta n(t) = n_2 I(t) + \int_0^\infty R(t') I(t-t') dt'$, where $R(t)$ is a response function that depends on properties of the rotational levels and the nuclear spin statistics \cite{nibbering_determination_1997,chen_single-shot_2007}.
In N$_2$ the rotational response peaks about 80 fs after the pump pulse arrives, as can be seen in Fig.~\ref{nitrogen}a.
The ratio of the instantaneous Kerr effect for parallel to perpendicular polarization is 3:1 in an isotropic medium.
For the rotational component, the ratio is $2:-1$.
The different symmetry properties allow the clean separation of the two contributions, as shown in Fig.~\ref{nitrogen}c.
Previous measurements using this technique \cite{chen_measurement_2007,chen_single-shot_2007} were unable to resolve the two contributions, but here we can owing to the shorter pump pulse.
To our knowledge, this is the first direct observation of the relative contributions of the instantaneous and rotational components of the Kerr effect in N$_2$.
At higher intensities plasma is observed, as shown in plots of $\Delta \phi(\xpump,t)$ in Fig.~\ref{nitrogen}de.

In summary, we have performed direct measurements of the optical Kerr effect in Ar and N$_2$ using single-shot supercontinuum spectral interferometry.
In N$_2$, we are able to distinguish between instantaneous and rotational components of the nonlinearity, and the polarization dependence is consistent with theory.
We observe the usual optical Kerr effect, linear in the intensity, as well as the onset of plasma, but no higher-order instantaneous nonlinearities effecting either saturation or negative response.
The fact that the pump-induced response appears to be linear in the intensity until the point of ionization is a reflection of the latter's extremely nonlinear onset.
At least for 38 fs pulses, there is no practical distinction between plasma and special atomic states with negative polarizability.
Finally, these results disprove the idea that higher-order instantaneous nonlinearities are important in nonlinear optics in gases at high intensities.
Our results strongly confirm the long-standing conceptual picture \cite{couairon_femtosecond_2007} that short pulse filamentation in gases arises from the interplay between nonlinear self-focusing from bound electron nonlinearities and defocusing due to plasma generation.

J.K.W. thanks the Joint Quantum Institute for support.
We thank S. Varma for helpful discussions.
This research was supported by the National Science Foundation, the U.S. Department of Energy, the Office of Naval Research, and the Lockheed Martin Corporation.

\appendix

\section{Appendix}
\begin{figure}
\center{\includegraphics[width=8.5cm]{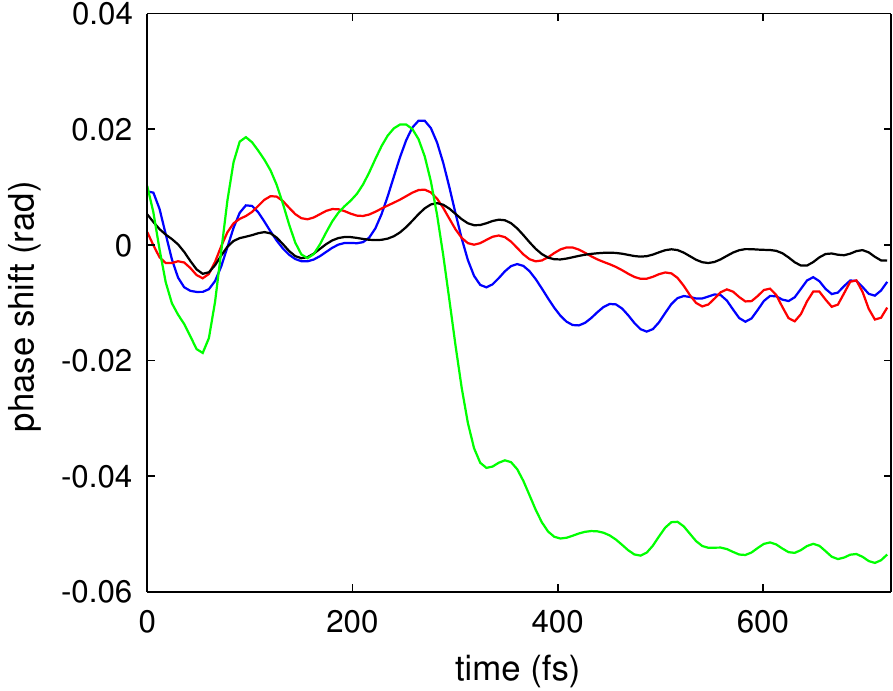}}
\caption{Experimental data for Ar with probe and reference spectra centered at 800 nm and $\Epump \perp \Eprobe$.
The measured phase shift $\Delta \phi(\xpump,t)$ is shown, extracted after averaging 1000 interferograms, at four pump intensities: 25 TW/cm$^2$ (black), 50 TW/cm$^2$ (red) , 75 TW/cm$^2$ (blue), and 100 TW/cm$^2$ (green).
}
\label{degen}
\end{figure}

Figure \ref{degen} shows experimental data in Ar using probe and reference pulses whose power spectra are identical to the pump pulse.
In this case the probe and reference pulses are not supercontinua, but rather chirped versions of the pump pulse.
To reject most of the pump light before the spectrometer, the pump polarization is perpendicular to the probe polarization, and a polarizer is inserted before the spectrometer.
The signal from the windows is larger with a probe centered at 800 nm because the dichroic mirror that is used as an exit window when the probe is supercontinuum centered at 680 nm is replaced with a fused silica window.
The window contribution has been subtracted in the data shown in Fig.~\ref{degen}.

Due to the leakage of pump light through the polarizer due to its finite extinction ratio, a small fraction of the pump light reaches the CCD camera in the spectrometer.
We attribute the oscillations in this data to the interference of this residual pump light with the reference pulse; as with the supercontinuum generated by the pump, this interference decreases when interferograms are averaged because the pump and probe arms are not phase stabilized.
In this data, 1000 interferograms were averaged before the phase shift was extracted.
The time resolution with the 800 nm probe is $\sim 40$ fs because of the narrower bandwidth.
As with the data taken with a supercontinuum probe, a monotonically increasing, positive instantaneous phase shift is seen up to the ionization threshold.
Note that coherent scattering from a plasma grating caused by interference between the pump and probe \cite{wahlstrand_effect_2011} is not present in this data because the pump polarization is perpendicular to the probe polarization.


\begin{thebibliography}{30}
\expandafter\ifx\csname natexlab\endcsname\relax\def\natexlab#1{#1}\fi
\expandafter\ifx\csname bibnamefont\endcsname\relax
  \def\bibnamefont#1{#1}\fi
\expandafter\ifx\csname bibfnamefont\endcsname\relax
  \def\bibfnamefont#1{#1}\fi
\expandafter\ifx\csname citenamefont\endcsname\relax
  \def\citenamefont#1{#1}\fi
\expandafter\ifx\csname url\endcsname\relax
  \def\url#1{\texttt{#1}}\fi
\expandafter\ifx\csname urlprefix\endcsname\relax\def\urlprefix{URL }\fi
\providecommand{\bibinfo}[2]{#2}
\providecommand{\eprint}[2][]{\url{#2}}

\bibitem{agrawal_nonlinear_2006}
\bibinfo{author}{\bibfnamefont{G.~P.} \bibnamefont{Agrawal}},
  \emph{\bibinfo{title}{Nonlinear Fiber Optics}} (\bibinfo{publisher}{Academic
  Press}, \bibinfo{year}{2006}), \bibinfo{edition}{4th} ed.

\bibitem{haus_mode-locking_2000}
\bibinfo{author}{\bibfnamefont{H.~A.} \bibnamefont{Haus}},
  \bibinfo{journal}{IEEE J. Sel. Topics Quantum Electron.}
  \textbf{\bibinfo{volume}{6}}, \bibinfo{pages}{1173} (\bibinfo{year}{2000}).

\bibitem{couairon_femtosecond_2007}
\bibinfo{author}{\bibfnamefont{A.}~\bibnamefont{Couairon}} \bibnamefont{and}
  \bibinfo{author}{\bibfnamefont{A.}~\bibnamefont{Mysyrowicz}},
  \bibinfo{journal}{Phys. Rep.} \textbf{\bibinfo{volume}{441}},
  \bibinfo{pages}{47} (\bibinfo{year}{2007}).

\bibitem{loriot_measurement_2009}
\bibinfo{author}{\bibfnamefont{V.}~\bibnamefont{Loriot}},
  \bibinfo{author}{\bibfnamefont{E.}~\bibnamefont{Hertz}},
  \bibinfo{author}{\bibfnamefont{O.}~\bibnamefont{Faucher}}, \bibnamefont{and}
  \bibinfo{author}{\bibfnamefont{B.}~\bibnamefont{Lavorel}},
  \bibinfo{journal}{Opt. Express} \textbf{\bibinfo{volume}{17}},
  \bibinfo{pages}{13429} (\bibinfo{year}{2009}).

\bibitem{bejot_higher-order_2010}
\bibinfo{author}{\bibfnamefont{P.}~\bibnamefont{B\'{e}jot}},
  \bibinfo{author}{\bibfnamefont{J.}~\bibnamefont{Kasparian}},
  \bibinfo{author}{\bibfnamefont{S.}~\bibnamefont{Henin}},
  \bibinfo{author}{\bibfnamefont{V.}~\bibnamefont{Loriot}},
  \bibinfo{author}{\bibfnamefont{T.}~\bibnamefont{Vieillard}},
  \bibinfo{author}{\bibfnamefont{E.}~\bibnamefont{Hertz}},
  \bibinfo{author}{\bibfnamefont{O.}~\bibnamefont{Faucher}},
  \bibinfo{author}{\bibfnamefont{B.}~\bibnamefont{Lavorel}}, \bibnamefont{and}
  \bibinfo{author}{\bibfnamefont{J.-P.}~\bibnamefont{Wolf}},
  \bibinfo{journal}{Phys. Rev. Lett.} \textbf{\bibinfo{volume}{104}},
  \bibinfo{pages}{103903} (\bibinfo{year}{2010}).

\bibitem{novoa_fermionic_2010}
\bibinfo{author}{\bibfnamefont{D.}~\bibnamefont{Novoa}},
  \bibinfo{author}{\bibfnamefont{H.}~\bibnamefont{Michinel}}, \bibnamefont{and}
  \bibinfo{author}{\bibfnamefont{D.}~\bibnamefont{Tommasini}},
  \bibinfo{journal}{Phys. Rev. Lett.} \textbf{\bibinfo{volume}{105}},
  \bibinfo{pages}{203904} (\bibinfo{year}{2010}).

\bibitem{ettoumi_generalized_2010}
W. Ettoumi, Y. Petit, J. Kasparian, and J.-P. Wolf, Opt. Express \textbf{18}, 6613 (2010).

\bibitem{kasparian_arbitrary-order_2010}
J. Kasparian, P. B\'ejot, and J.-P. Wolf, Opt. Lett. \textbf{35}, 2795 (2010).

\bibitem[{\citenamefont{Kolesik
  et~al.}(2010{\natexlab{b}})\citenamefont{Kolesik, Wright, and
  Moloney}}]{kolesik_femtosecond_2010}
\bibinfo{author}{\bibfnamefont{M.}~\bibnamefont{Kolesik}},
  \bibinfo{author}{\bibfnamefont{E.~M.} \bibnamefont{Wright}},
  \bibnamefont{and} \bibinfo{author}{\bibfnamefont{J.~V.}
  \bibnamefont{Moloney}}, \bibinfo{journal}{Opt. Lett.}
  \textbf{\bibinfo{volume}{35}}, \bibinfo{pages}{2550}
  (\bibinfo{year}{2010}{\natexlab{b}}).

\bibitem[{\citenamefont{B?jot et~al.}(2011)\citenamefont{B?jot, Hertz, Lavorel,
  Kasparian, Wolf, and Faucher}}]{bjot_higher-order_2011}
\bibinfo{author}{\bibfnamefont{P.}~\bibnamefont{B\'ejot}},
  \bibinfo{author}{\bibfnamefont{E.}~\bibnamefont{Hertz}},
  \bibinfo{author}{\bibfnamefont{B.}~\bibnamefont{Lavorel}},
  \bibinfo{author}{\bibfnamefont{J.}~\bibnamefont{Kasparian}},
  \bibinfo{author}{\bibfnamefont{J.-P.}~\bibnamefont{Wolf}}, \bibnamefont{and}
  \bibinfo{author}{\bibfnamefont{O.}~\bibnamefont{Faucher}},
  \bibinfo{journal}{Opt. Lett.} \textbf{\bibinfo{volume}{36}},
  \bibinfo{pages}{828} (\bibinfo{year}{2011}).

\bibitem{ariunbold_2011}
G. O. Ariunbold, P. Polynkin, and J. V. Moloney, arXiv:1106.5511 (unpublished).

\bibitem{kolesik_higher-order_2010}
\bibinfo{author}{\bibfnamefont{M.}~\bibnamefont{Kolesik}},
  \bibinfo{author}{\bibfnamefont{D.}~\bibnamefont{Mirell}},
  \bibinfo{author}{\bibfnamefont{J.}~\bibnamefont{Diels}}, \bibnamefont{and}
  \bibinfo{author}{\bibfnamefont{J.~V.} \bibnamefont{Moloney}},
  \bibinfo{journal}{Opt. Lett.} \textbf{\bibinfo{volume}{35}},
  \bibinfo{pages}{3685} (\bibinfo{year}{2010}{\natexlab{a}}).

\bibitem{chen_direct_2010}
\bibinfo{author}{\bibfnamefont{Y.-H.}~\bibnamefont{Chen}},
  \bibinfo{author}{\bibfnamefont{S.}~\bibnamefont{Varma}},
  \bibinfo{author}{\bibfnamefont{T.~M.} \bibnamefont{Antonsen}},
  \bibnamefont{and} \bibinfo{author}{\bibfnamefont{H.~M.}
  \bibnamefont{Milchberg}},
  \bibinfo{journal}{Phys. Rev. Lett.}
  \textbf{\bibinfo{volume}{105}}, \bibinfo{pages}{215005}
  (\bibinfo{year}{2010}).

\bibitem[{\citenamefont{Kosareva et~al.}(2011)\citenamefont{Kosareva, Daigle,
  Panov, Wang, Hosseini, Yuan, Roy, Makarov, and Chin}}]{kosareva_arrest_2011}
\bibinfo{author}{\bibfnamefont{O.}~\bibnamefont{Kosareva}},
  \bibinfo{author}{\bibfnamefont{J.}~\bibnamefont{Daigle}},
  \bibinfo{author}{\bibfnamefont{N.}~\bibnamefont{Panov}},
  \bibinfo{author}{\bibfnamefont{T.}~\bibnamefont{Wang}},
  \bibinfo{author}{\bibfnamefont{S.}~\bibnamefont{Hosseini}},
  \bibinfo{author}{\bibfnamefont{S.}~\bibnamefont{Yuan}},
  \bibinfo{author}{\bibfnamefont{G.}~\bibnamefont{Roy}},
  \bibinfo{author}{\bibfnamefont{V.}~\bibnamefont{Makarov}}, \bibnamefont{and}
  \bibinfo{author}{\bibfnamefont{S.~L.} \bibnamefont{Chin}},
  \bibinfo{journal}{Opt. Lett.} \textbf{\bibinfo{volume}{36}},
  \bibinfo{pages}{1035} (\bibinfo{year}{2011}).

\bibitem[{\citenamefont{Polynkin et~al.}(2011)\citenamefont{Polynkin, Kolesik,
  Wright, and Moloney}}]{polynkin_experimental_2011}
\bibinfo{author}{\bibfnamefont{P.}~\bibnamefont{Polynkin}},
  \bibinfo{author}{\bibfnamefont{M.}~\bibnamefont{Kolesik}},
  \bibinfo{author}{\bibfnamefont{E.~M.} \bibnamefont{Wright}},
  \bibnamefont{and} \bibinfo{author}{\bibfnamefont{J.~V.}
  \bibnamefont{Moloney}},
  \bibinfo{journal}{Phys. Rev. Lett.}
  \textbf{\bibinfo{volume}{106}}, \bibinfo{pages}{153902}
  (\bibinfo{year}{2011}).

\bibitem{bejot_transition_2011}
\bibinfo{author}{\bibfnamefont{P.}~\bibnamefont{B\'ejot}},
E. Hertz, J. Kasparian, B. Lavorel, J.-P. Wolf, and O. Faucher,
  \bibinfo{journal}{Phys. Rev. Lett.} \textbf{106}, 243902 (2011).

\bibitem[{\citenamefont{Teleki et~al.}(2010)\citenamefont{Teleki, Wright, and
  Kolesik}}]{teleki_microscopic_2010}
\bibinfo{author}{\bibfnamefont{A.}~\bibnamefont{Teleki}},
  \bibinfo{author}{\bibfnamefont{E.~M.} \bibnamefont{Wright}},
  \bibnamefont{and} \bibinfo{author}{\bibfnamefont{M.}~\bibnamefont{Kolesik}},
  \bibinfo{journal}{Phys. Rev. A} \textbf{\bibinfo{volume}{82}},
  \bibinfo{pages}{065801} (\bibinfo{year}{2010}).

\bibitem[{\citenamefont{Br\'{e}e et~al.}(2011)\citenamefont{Br\'{e}e, Demircan,
  and Steinmeyer}}]{bree_saturation_2011}
\bibinfo{author}{\bibfnamefont{C.}~\bibnamefont{Br\'{e}e}},
  \bibinfo{author}{\bibfnamefont{A.}~\bibnamefont{Demircan}}, \bibnamefont{and}
  \bibinfo{author}{\bibfnamefont{G.}~\bibnamefont{Steinmeyer}},
  \bibinfo{journal}{Phys. Rev. Lett.} \textbf{\bibinfo{volume}{106}},
  \bibinfo{pages}{183902} (\bibinfo{year}{2011}).

\bibitem[{\citenamefont{Kim et~al.}(2002{\natexlab{a}})\citenamefont{Kim,
  Alexeev, and Milchberg}}]{kim_single-shot_2002-1}
\bibinfo{author}{\bibfnamefont{K.~Y.} \bibnamefont{Kim}},
  \bibinfo{author}{\bibfnamefont{I.}~\bibnamefont{Alexeev}}, \bibnamefont{and}
  \bibinfo{author}{\bibfnamefont{H.~M.} \bibnamefont{Milchberg}},
  \bibinfo{journal}{Appl. Phys. Lett.} \textbf{\bibinfo{volume}{81}},
  \bibinfo{pages}{4124} (\bibinfo{year}{2002}{\natexlab{a}}).

\bibitem[{\citenamefont{Stegeman et~al.}(2011)\citenamefont{Stegeman,
  Papazoglou, Boyd, and Tzortzakis}}]{stegeman_nonlinear_2011}
Going from a measurement of the transient birefringence to Kerr coefficients requires careful accounting of all relevant tensor components, as discussed in
\bibinfo{author}{\bibfnamefont{G.}~\bibnamefont{Stegeman}},
  \bibinfo{author}{\bibfnamefont{D.~G.} \bibnamefont{Papazoglou}},
  \bibinfo{author}{\bibfnamefont{R.}~\bibnamefont{Boyd}}, \bibnamefont{and}
  \bibinfo{author}{\bibfnamefont{S.}~\bibnamefont{Tzortzakis}},
  \bibinfo{journal}{Opt. Express} \textbf{\bibinfo{volume}{19}},
  \bibinfo{pages}{6387} (\bibinfo{year}{2011}).

\bibitem[{\citenamefont{Chen et~al.}(2007{\natexlab{a}})\citenamefont{Chen,
  Varma, Alexeev, and Milchberg}}]{chen_measurement_2007}
\bibinfo{author}{\bibfnamefont{Y.-H.}~\bibnamefont{Chen}},
  \bibinfo{author}{\bibfnamefont{S.}~\bibnamefont{Varma}},
  \bibinfo{author}{\bibfnamefont{I.}~\bibnamefont{Alexeev}}, \bibnamefont{and}
  \bibinfo{author}{\bibfnamefont{H.}~\bibnamefont{Milchberg}},
  \bibinfo{journal}{Opt. Express} \textbf{\bibinfo{volume}{15}},
  \bibinfo{pages}{7458} (\bibinfo{year}{2007}{\natexlab{a}}).

\bibitem[{\citenamefont{Kim et~al.}(2002{\natexlab{b}})\citenamefont{Kim,
  Alexeev, and Milchberg}}]{kim_single-shot_2002}
The distorting effects of using an interaction length in spectral interferometry much longer than the Rayleigh range are discussed and simulated in \bibinfo{author}{\bibfnamefont{K. Y.}~\bibnamefont{Kim}},
  \bibinfo{author}{\bibfnamefont{I.}~\bibnamefont{Alexeev}}, \bibnamefont{and}
  \bibinfo{author}{\bibfnamefont{H. M.}~\bibnamefont{Milchberg}},
  \bibinfo{journal}{Opt. Express} \textbf{\bibinfo{volume}{10}},
  \bibinfo{pages}{1563} (\bibinfo{year}{2002}{\natexlab{b}}).

\bibitem[{\citenamefont{Marceau et~al.}(2010)\citenamefont{Marceau,
  Ramakrishna, Génier, Wang, Chen, Théberge, Châteauneuf, Dubois, Seideman,
  and Chin}}]{marceau_femtosecond_2010}
\bibinfo{author}{\bibfnamefont{C.}~\bibnamefont{Marceau}},
  \bibinfo{author}{\bibfnamefont{S.}~\bibnamefont{Ramakrishna}},
  \bibinfo{author}{\bibfnamefont{S.}~\bibnamefont{G\'enier}},
  \bibinfo{author}{\bibfnamefont{T.}~\bibnamefont{Wang}},
  \bibinfo{author}{\bibfnamefont{Y.}~\bibnamefont{Chen}},
  \bibinfo{author}{\bibfnamefont{F.}~\bibnamefont{Th\'eberge}},
  \bibinfo{author}{\bibfnamefont{M.}~\bibnamefont{Ch\^ateauneuf}},
  \bibinfo{author}{\bibfnamefont{J.}~\bibnamefont{Dubois}},
  \bibinfo{author}{\bibfnamefont{T.}~\bibnamefont{Seideman}}, \bibnamefont{and}
  \bibinfo{author}{\bibfnamefont{S.~L.} \bibnamefont{Chin}},
  \bibinfo{journal}{Opt. Commun.} \textbf{\bibinfo{volume}{283}},
  \bibinfo{pages}{2732} (\bibinfo{year}{2010}).

\bibitem[{\citenamefont{Lehmeier et~al.}(1985)\citenamefont{Lehmeier,
  Leupacher, and Penzkofer}}]{lehmeier_nonresonant_1985}
\bibinfo{author}{\bibfnamefont{H.}~\bibnamefont{Lehmeier}},
  \bibinfo{author}{\bibfnamefont{W.}~\bibnamefont{Leupacher}},
  \bibnamefont{and}
  \bibinfo{author}{\bibfnamefont{A.}~\bibnamefont{Penzkofer}},
  \bibinfo{journal}{Opt. Commun.} \textbf{\bibinfo{volume}{56}},
  \bibinfo{pages}{67} (\bibinfo{year}{1985}).

\bibitem{wahlstrand_effect_2011}
J. K. Wahlstrand and H. M. Milchberg, arXiv:1107.2830 (unpublished).

\bibitem[{\citenamefont{Larochelle
  et~al.}(1998{\natexlab{a}})\citenamefont{Larochelle, Talebpour, and
  Chin}}]{larochelle_non-sequential_1998}
\bibinfo{author}{\bibfnamefont{S.}~\bibnamefont{Larochelle}},
  \bibinfo{author}{\bibfnamefont{A.}~\bibnamefont{Talebpour}},
  \bibnamefont{and} \bibinfo{author}{\bibfnamefont{S.~L.} \bibnamefont{Chin}},
  \bibinfo{journal}{J. Phys. B: At. Mol. Opt. Phys.} \textbf{\bibinfo{volume}{31}}, \bibinfo{pages}{1201}
  (\bibinfo{year}{1998}{\natexlab{a}});
\bibinfo{author}{\bibfnamefont{S.~F.~J.} \bibnamefont{Larochelle}},
  \bibinfo{author}{\bibfnamefont{A.}~\bibnamefont{Talebpour}},
  \bibnamefont{and} \bibinfo{author}{\bibfnamefont{S.~L.} \bibnamefont{Chin}},
  \bibinfo{journal}{J. Phys. B: At. Mol. Opt. Phys.} \textbf{\bibinfo{volume}{31}}, \bibinfo{pages}{1215}
  (\bibinfo{year}{1998}{\natexlab{b}}).

\bibitem[{\citenamefont{Kosareva et~al.}(1997)\citenamefont{Kosareva, Kandidov,
  Brodeur, and Chin}}]{kosareva_article_1997}
\bibinfo{author}{\bibfnamefont{O.~G.} \bibnamefont{Kosareva}},
  \bibinfo{author}{\bibfnamefont{V.~P.} \bibnamefont{Kandidov}},
  \bibinfo{author}{\bibfnamefont{A.}~\bibnamefont{Brodeur}}, \bibnamefont{and}
  \bibinfo{author}{\bibfnamefont{S.~L.} \bibnamefont{Chin}},
  \bibinfo{journal}{J. Nonlinear Opt. Phys. Mater.}
  \textbf{\bibinfo{volume}{6}}, \bibinfo{pages}{485} (\bibinfo{year}{1997}).

\bibitem{ammosov}
M. V. Ammosov, N. B. Delone and V. P. Krainov. Sov. Phys. JETP \textbf{64}, 1191 (1986).

\bibitem[{\citenamefont{Nibbering et~al.}(1997)\citenamefont{Nibbering,
  Grillon, Franco, Prade, and Mysyrowicz}}]{nibbering_determination_1997}
\bibinfo{author}{\bibfnamefont{E.~T.~J.} \bibnamefont{Nibbering}},
  \bibinfo{author}{\bibfnamefont{G.}~\bibnamefont{Grillon}},
  \bibinfo{author}{\bibfnamefont{M.~A.} \bibnamefont{Franco}},
  \bibinfo{author}{\bibfnamefont{B.~S.} \bibnamefont{Prade}}, \bibnamefont{and}
  \bibinfo{author}{\bibfnamefont{A.}~\bibnamefont{Mysyrowicz}},
  \bibinfo{journal}{J. Opt. Soc. Am. B}
  \textbf{\bibinfo{volume}{14}}, \bibinfo{pages}{650} (\bibinfo{year}{1997}).

\bibitem[{\citenamefont{Chen et~al.}(2007{\natexlab{b}})\citenamefont{Chen,
  Varma, York, and Milchberg}}]{chen_single-shot_2007}
\bibinfo{author}{\bibfnamefont{Y.-H.}~\bibnamefont{Chen}},
  \bibinfo{author}{\bibfnamefont{S.}~\bibnamefont{Varma}},
  \bibinfo{author}{\bibfnamefont{A.}~\bibnamefont{York}}, \bibnamefont{and}
  \bibinfo{author}{\bibfnamefont{H.~M.} \bibnamefont{Milchberg}},
  \bibinfo{journal}{Opt. Express} \textbf{\bibinfo{volume}{15}},
  \bibinfo{pages}{11341} (\bibinfo{year}{2007}{\natexlab{b}}).

\end{thebibliography}
\end{document}